\documentclass[aip,graphicx]{revtex4-1}

\draft % marks overfull lines with a black rule on the right
\usepackage{graphicx}
\usepackage{xcolor}
\usepackage[version=4,arrows=pgf-filled,
textfontname=sffamily,
mathfontname=mathsf]{mhchem}

\begin{document}

%\preprint{}

\title{Polyelectrolyte adsorption at the solid-liquid interface favors receding contact line instability} %Title of paper



\author{Léa Delance}
%\thanks{}
\affiliation{Max Planck Institute for Polymer Research, Ackermannweg 10, 55128 Mainz, Germany}
\author{Diego Díaz}
\affiliation{KTH Royal Institute of Technology, SE-10044 Stockholm, Sweden.}
\author{Arivazhagan G. Balasubramanian}
\affiliation{KTH Royal Institute of Technology, SE-10044 Stockholm, Sweden.}
\author{Outi Tammisola}
\affiliation{KTH Royal Institute of Technology, SE-10044 Stockholm, Sweden.}
\author{Kaloian Koynov}
\affiliation{Max Planck Institute for Polymer Research, Ackermannweg 10, 55128 Mainz, Germany}
\author{Hans-Jürgen Butt}
\affiliation{Max Planck Institute for Polymer Research, Ackermannweg 10, 55128 Mainz, Germany}

\date{\today}

\begin{abstract}
Controlling the motion of non-Newtonian drops on surfaces is crucial for applications ranging from inkjet printing to biomedical devices and food processing. While the macroscopic behavior of viscoelastic drops sliding on tilted hydrophobic surfaces has been characterized—showing reduced velocities and elongation compared to Newtonian fluids—the underlying microscopic mechanisms remain poorly understood.
To address this gap, we developed a high-speed, high-resolution reflection microscope that enables direct visualization of the contact line of sliding drops. We used water-soluble polyelectrolyte solutions based on polyacrylamide and let drops sliding on hydrophobic substrates composed of Teflon AF (amorphous fluoropolymer)- and PDMS (Polydimethylsiloxane)-coated glass slides. The substrate tilting angle was varied between 20° and 45°. We reveal how viscoelasticity influences the dynamics of the receding contact line and drop motion. Our experiments demonstrate that viscoelasticity can destabilize the receding contact line, triggering filament formation. This instability previously observed in the coating of thin viscoelastic films, is reported here for sliding drops, for the first time on smooth surfaces.
We further highlight the critical role of polymer charge in this process: while cationic and non-ionic polymers promote filament formation, anionic polymers do not, a difference we attribute to the distinct wetting properties of the solutions. In conclusion, we clarify the interplay between rheology, surface interactions, and drop dynamics.

\end{abstract}


\maketitle %\maketitle must follow title, authors, abstract and \pacs

\section{Introduction}

Viscous fingering or Saffman-Taylor instability \cite{saffman_penetration_1997} is a well studied instability that arises when a less viscous fluid displaces a more viscous fluid, leading to the formation of "fingers". Fingering instabilities also occur at contact lines. For example, at advancing contact lines, when a fluid front is flowing down a tilted plate, regions with larger height flow faster leading to the formation of fingers\cite{huppert_flow_1982}. This instability is primarily driven by macroscopic fluid dynamics, with contact line dynamics playing a minor role\cite{brenner_instability_1993}. More recently, fingering instabilities have also been observed at receding contact lines in the context of non-Newtonian fluids, particularly viscoelastic fluids composed of an aqueous solution of high-molecular weight polymers\cite{deblais_taming_2016}. The origin of this instability differs from that at advancing contact lines: it arises from the interplay between Van der Waals and capillary forces\cite{sharma_newtonian_2025}. This results in a distinct dependency on contact line velocity. Unlike advancing contact lines, the wavelength of fingers forming at the receding contact line is found to increase with contact line velocity. However, this study does not account for fluid elasticity and does not explain the evolution of filament length with contact line velocity.

A drop sliding across a tilted plate provides a system that allows for the study of both advancing and receding contact lines. The literature on viscoelastic sliding drops is limited and primarily focuses on macroscopic features such as velocity and shape. Varagnolo \textit{et al.}\cite{varagnolo_stretching_2017} showed that sliding viscoelastic drops are elongated compared to Newtonian fluids, with the elongation depending on polymer flexibility and molecular weight. Xu \textit{et al.}\cite{xu_sliding_2016, xu_viscoelastic_2018} demonstrated the deposition of filaments at the rear of drops on superhydrophobic surfaces consisting of pillars. They linked filament formation to the interplay between gravity, viscosity, and contact angle hysteresis. However, many aspects remain unclear: How do the physico-chemical properties of sliding drops affect contact line instabilities? Are polymers deposited on the surface during contact line motion? Understanding these questions is valuable for applications that rely on viscoelastic fluids of high molecular weight, including coating processes and transport of biological materials.

In this work, we explore the microscopic deformation of the receding contact lines of viscoelastic drops sliding across a hydrophobic surface. It turned out that viscoelastic sliding drops exhibit receding contact line instabilities leading to the formation of filaments. We demonstrate how the physico-chemical properties of the drop influence the onset of instability and filament formation. We study polyacrylamide (repeat unit -CH$_2$CH(CONH$_2$)-)
%$\ce{(C3H5NO)}_n$) This is only a suggestion to make the structure more clear OK
and two polyacrylamide-based polyelectrolytes. In these, a certain fraction of acrylamide units have been replaced by units bearing a non-zero charge in solution. When in solution, the anionic and cationic polyelectrolytes release their counter-ions and get charged. Polyacrylamide is generally considered to be non-ionic. As shown in \cite{zimmermann_electrokinetic_2001,https://doi.org/10.1002/elps.200700734,PREOCANIN2012120}, zeta-potential measurements indicate that Teflon AF (amorphous fluoropolymer) surfaces acquire a negative charge and spontaneoulsy form an electric double layer when exposed to water at neutral pH. 
%The surface is surrounded by the electric double layer. I tried to avoid the term "surrounded" in the context of a surface. OK
We show that the occurrence of instability depends on the charge of the polymer in solution and is due to polymer/surface interactions. Cationic and non-ionic polymers are more prone to develop long filaments while anionic only show small deformation of the receding contact line. Finally, we demonstrate that these filaments lead to the deposition of polymer on the surface.

\section{Material and methods}

\subsection{Polymer solutions}

Three polymers were used: polyacrylamide (FLOPAM FA 920 SH), non-ionic; a copolymer of acrylamide and acrylic acid (SNF, FLOPAM AN934 SH), anionic; and a copolymer of acrylamide and chloro-methylated monomer (SNF, FLOPAM FO 4290 SH). All polymers have high molecular weight($5-15\times 10^{-6}$ g/mol) and are water-soluble. Solutions at 0.025\% weight fraction were prepared by dissolving the polymer in milliQ water and stirring for two days until complete dissolution.

The shear-viscosity of the solutions was determined using a DHR-3 rheometer (TA Instruments) with a cone-plate geometry (2° angle). All fluids showed shear-thinning behavior, i.e, its viscosity decreased under applied shear rate (See supporting information, SI). The temperature was maintained at $25\pm 1 $~°C. The relaxation time $\tau$ of the solutions was determined using the Dripping onto surface method~\cite{dinic2015extensional,dinic2017pinch, dinic2019macromolecular, robertson2022evaporation}. In this method, a drop was deposited on a surface and the thinning of the filament (liquid bridge) formed during deposition was recorded by a high speed camera at 5000 fps. The diameter of the filament was evaluated by ImageJ Software over time. The relaxation time $\tau$ 
% use a different symbol, not lambda. Lambda is used later for a length scale, which is good. maybe use tau. OK
was determined by fitting $\exp{(-t/3\tau)}$~\cite{mckinley2005visco} to the exponential thinning decay of the filament. The aspect ratio between the distance from nozzle to substrate (glass slide) $H$ and nozzle radius $r_0$ was $H/r_0=3$.

The surface tension of the solutions was measured using the pendant drop method and was $71\pm 1$~mN/m. 

\subsection{Surface preparation}

The substrate consisted of glass slides with a thickness of 0.17 mm coated on the backside with a conductive ITO layer (Präzisions Glas \& Optik GmbH). Two different hydrophobic surface coatings were pepared. Teflon AF-coated slides were fabricated by dip-coating the glass slides with a 60 nm layer of Teflon AF from a 1 wt\% Teflon AF  at a withdrawal speed of 1 cm/min, followed by annealing at 160°C under vacuum for 24\;h. PDMS (Polydimethylsiloxane)-coated slides were prepared using a grafting-to method.\cite{https://doi.org/10.1002/adma.202311470} The glass slides were drop cast with polydimethylsiloxane (PDMS) with a molecular weight of 6000 g/mol (Sigma Aldrich), placed into an oven at 100°C for 24\;h, and subsequently rinsed with toluene and ethanol to remove free PDMS.

\subsection{Reflection microscopy of sliding drops}

\begin{figure}
    \centering
    \includegraphics[width=0.99\linewidth]{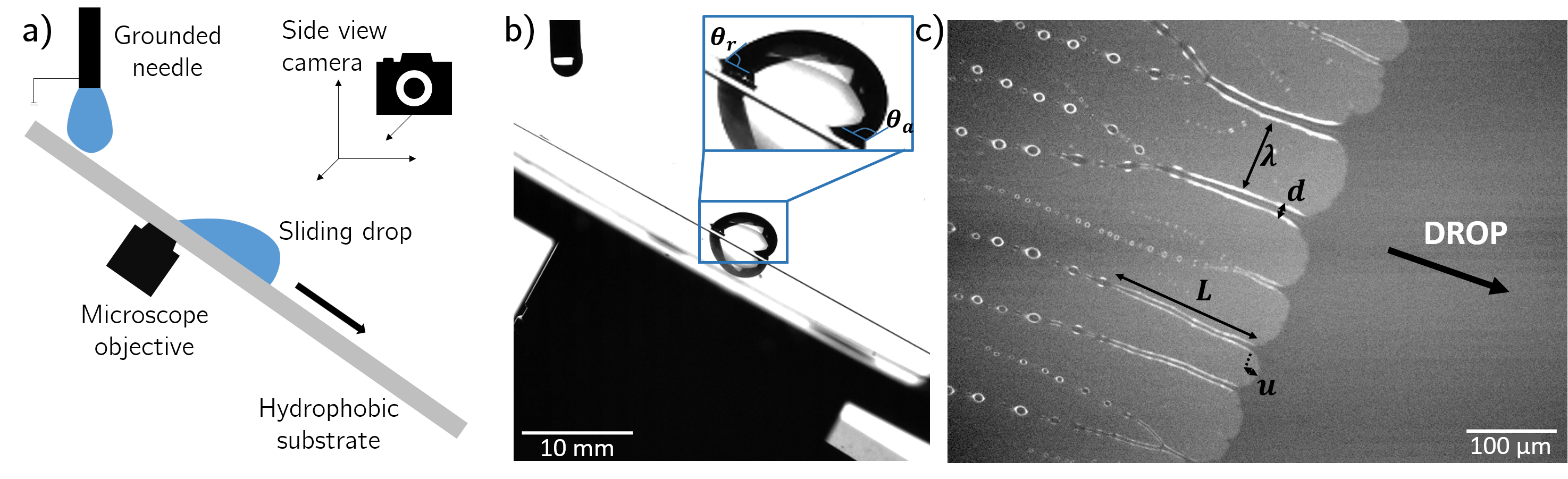}
    \caption{a) Schematics of the setup used for reflection microscopy of sliding drops. b-c) Typical images obtained respectively from the side-view and bottom-view camera.}
    \label{fig:setup}
\end{figure}

We performed bottom-view imaging of contact line dynamics during drop sliding (Fig.\ref{fig:setup}). An inverted epifluorescence microscope (Olympus IX83) was operated in reflection mode with a 20$\times$ objective (Olympus UCPLANFL N 20x) and a high-power LED (Thorlabs DC2200, 525 nm). The sample was placed on a grounded metallic stage modified using an objective inverter to allow for the observation of a tilted slide, as has been described in a previous article. \cite{https://doi.org/10.1002/dro2.70039} Images were acquired at 10 000 fps using a high-speed camera (Photron, Phantom TMX 7510).

Reflection microscopy was combined with side-view imaging of the sliding drop using a second high-speed camera (Photron Fastcam Mini AX10, 125 fps) and a telecentric lens (Edmund Optics CobaltTL 0.274x). This enabled measurement of the macroscopic drop velocity and the advancing and receding contact angles $\theta_a$ and $\theta_r$.

$45 \pm 3\;\mu$L drops were formed from a grounded needle and allowed to slide on the studied hydrophobic surface. The microscope objective field of view was positioned at a distance of 2 cm from the beginning of sliding. For each drop both bottom-view and side-view images were recorded simultaneously.

After sliding, we checked the deposition of polymer using Scanning Electron Microscopy (ZEISS Gemini SEM 560). To do so, samples were coated after sliding with a 2.5\;nm thick layer of platinum to avoid charging and were imaged with an accelerating voltage of 3\;kV. 
% Maybe shift the descrition of SEM to M & M. We should certainly also reprt the type of SEM. Which brand? OK

\section{Results and Discussion}

\begin{figure}
    \centering
    \includegraphics[width=0.99\linewidth]{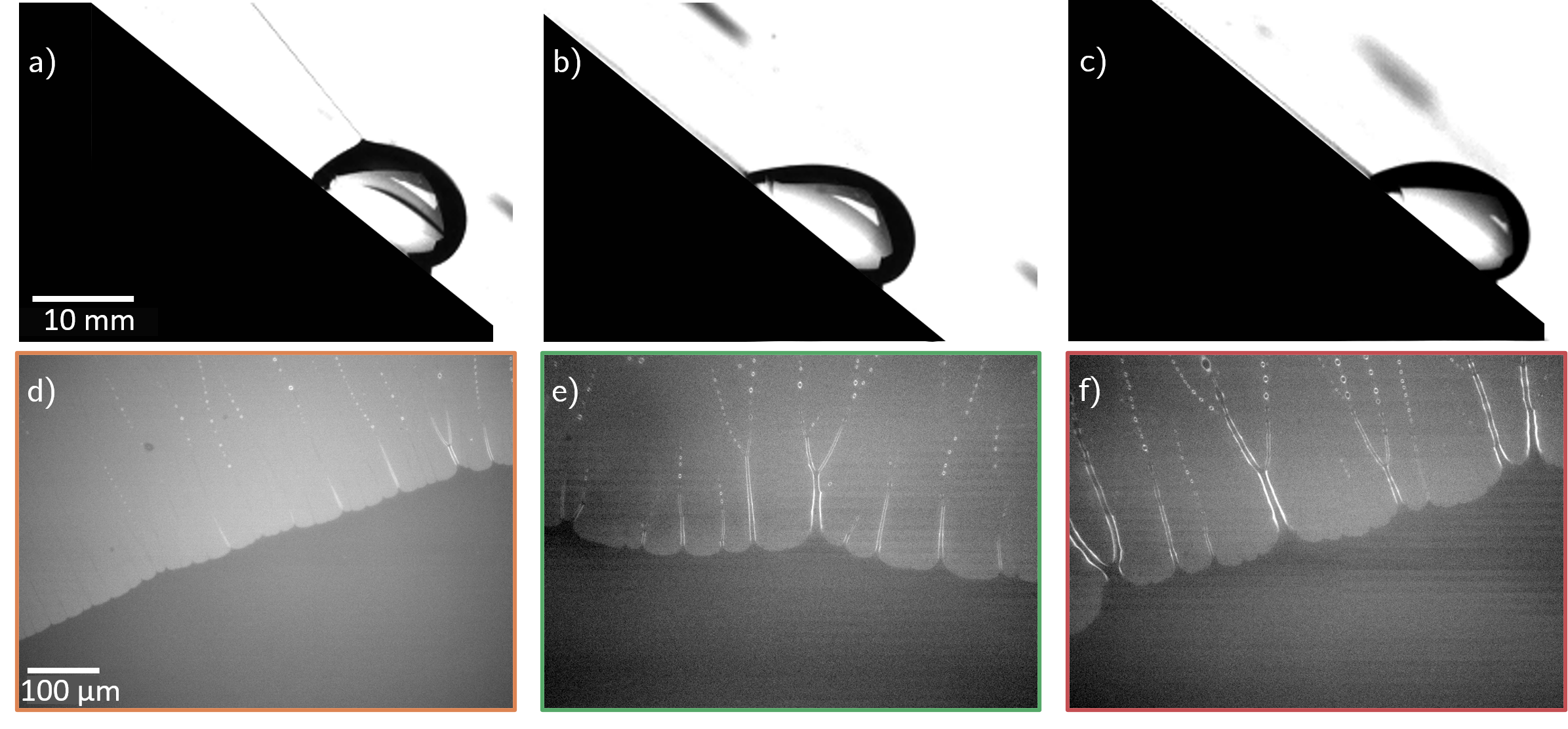}
    \caption{Side-view images of the drops and reflection microscope images of the receding contact line for a tilting angle of 40° for the anionic polymer (respectively a,d), the non-ionic polymer (b,e), and the cationic polymer (c,f). The dark part correspond to a liquid/solid interface while the brighter part correspond to the air/solid interface.}
    \label{fig:images_filaments}
\end{figure}

For all solutions, side view (Fig.\;\ref{fig:images_filaments}a-c) and reflection microscope videos (Fig.\;\ref{fig:images_filaments}d-f) showed that the the advancing contact line (not shown) is smooth at the scale of $\approx\!10\;\mu$m and exhibited no significant difference in shape from that of a pure water drop. In the following, we focus on the receding contact line. For the anionic polyelectrolyte (Fig.\;\ref{fig:images_filaments}d), we observed small deformations of the contact line extending $u\approx\! 5\; \mu$m in the direction of sliding. The protrusions are associated with the formation of thin filaments (d$\approx 2\;\mu$m thickness). 
% I suggest not to use the symbol "e" for diameter. e is e. Sometimes the unit charge. But not good for diameter. d would be good for diameter. OK
The protrusions are regularly spaced at $\lambda \approx\! 15\; \mu$m intervals but the formation of filaments is not continuous along the contact line: some protrusions did not result in filament formation. At lower tilt angles and thus lower speed, no filaments formed.

Cationic and non-ionic polymer drops (Fig.\;\ref{fig:images_filaments}e,f) exhibited larger deformation (with an extension of $u\approx\!20\;\mu$m) and the formation of thicker ($d\!\approx\!10\;\mu$m), and longer filaments ($L\!\approx\!100\;\mu$m). The filaments destabilized at the end forming regularly spaced microdroplets.  Filaments formed at every protrusion and were regularly spaced. 

\begin{figure}
    \centering
    \includegraphics[width=0.99\linewidth]{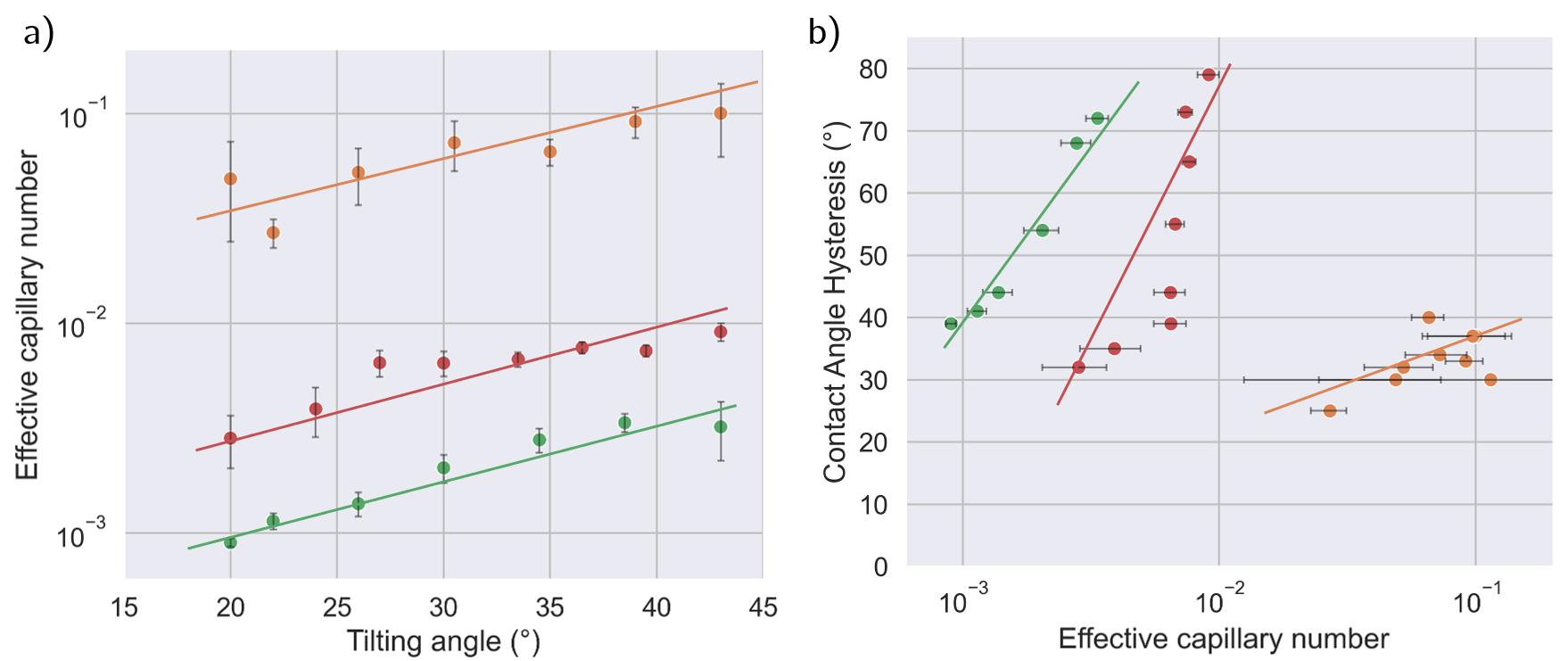}
    % Axis are missing!
    \caption{(a) Effective capillary number as a function of the tilting angle and (b) contact angle hysteresis as a function of the capillary number for the anionic (orange), non-ionic (green), and cationic (red) polymers. Error bars were computed based on the standard deviation of the averaged velocity.
    }
    \label{fig:velocity_CAH}
\end{figure}

Drops containing the anionic polymer moved faster than non-ionic and cationic polymer drops. After sliding 3\;cm the typical speeds were 30 to 60 mm/s for drops with dissolved anionic polymer. The non-ionic and the cationic polymer drops slid with a velocity of roughly 5 to 50 mm/s and 1 to 10 mm/s, respectively. While cationic and non-ionic polymer drops reached a steady state velocity, anionic polymer drops were still accelerating after a slide length of 3 cm. In the following, we report the average velocity measured after a slide length between 2.5 and 3.5\;cm. 

To explore the reasons for the observed differences in the macroscopic motion, we performed rheology measurements (see Supplementary Information, Fig. S1) and found that the zero-shear viscosity of the polymer solutions differs with several orders of magnitude and that the solutions are shear-thinning. The zero-shear viscosities ranged from 0.009 Pa\;s (non-ionic) to 0.4 Pa\;s (cationic) and 4.9 Pa\;s (anionic).
To account for these differences, we computed the effective capillary number $Ca=\eta_{eff}(\dot{\gamma})U / \sigma$, where $\eta$ and $\dot{\gamma}$ are the characteristic viscosity and shear rate, respectively. Here, the characteristic shear rate is $\dot{\gamma}=U/D_0$ with $D_0$ as the drop diameter. 
% Below we call this shear rate "bulk" shear rate. $\dot{\gamma}_{bulk}$ It should have the same name as below. OK
The average drop velocity $U$ was obtained from the side view images by measuring and dividing it by the corresponding sliding time. The viscosity $\eta$ was determined from rheology measurements matching the characteristic shear rate with the steady shear curves (see SI). The surface tension $\sigma$ is constant and indistinguishable from the one of pure water. We found that drops with anionic polymer have capillary numbers at least one order of magnitude higher than the non-ionic and cationic polymers (Fig.\;\ref{fig:velocity_CAH}a). Therefore, after scaling with viscosity, we find that non-ionic and cationic polymer drops are slowed down compared to the anionic polymer. We later discuss the effect of polymer charge on drop sliding.
% Shouldn't we give a reason why anionic polymer behaves so differently? Why does it? => Léa: polymer adsorption is discussed later in the paper. OK

Given that the driving force for sliding (gravity) remained unchanged, this slowdown indicates an increase in friction force. The friction force experienced by a sliding drop is described using the Furmidge-Kawasaki equation \cite{FURMIDGE1962309,KAWASAKI1960402,butt_drop_2025,mchale_adhesive_2025}: 
\begin{equation}
    F = w\sigma k(\cos \theta_r-\cos \theta_a) + F_{bulk},
\end{equation}
with $w$ the drop width, $k\!\approx\!0.8$ a geometrical factor,
% Here you can add "Adhesive forces in droplet kinetic friction on liquid-like surfaces" by Glen McHale, Sara Janahi, Hernán Barrio-Zhang, Yaofeng Wang, Jinju Chen, Gary G Wells, Rodrigo Ledesma-Aguilar, 2025 OK
% On page 7 we report the Furmidge equation. We should, however, be careful and not say that this is the friction for of a drop. In our case, viscous dissipation seems to be substantial. Furmidge only contains viscous dissipation very close to the contact line, but not in the bulk. So to get the full friction force we need to add bulk viscous dissipation. See eq. 7 in “Drop friction”, Nature Reviews Physics 2025, https://doi.org/10.1038/s42254-025-00841-5 . OK
and $\theta_{a/r}$ respectively the dynamic advancing and receding contact angles. $F_{bulk}$ is due to bulk viscous dissipation. Drop friction can therefore be quantified by measuring contact angle hysteresis, defined as $\Delta\theta = \theta_a - \theta_r$. We measured contact angles and $\Delta\theta$ for different tilting angles from side-view images at the moment the drop was observed with the microscope (Fig.\;\ref{fig:velocity_CAH}b). We measured the receding contact angle using the macroscopic shape of the drop and ignoring the tails observed for cationic and non-ionic polymer drops. Thus, the contact angles reported refer to the macroscale ($\le 10\;\mu$m) and can be considered apparent contact angles. For anionic polymer drops, the hysteresis remained stable (between 30 and 40°). In contrast, for non-ionic and cationic polymer drops, hysteresis increased and reached 70° for a tilt angle of 40°. This increase is primarily due to a decrease in the receding contact angle, while the advancing contact angle remains nearly constant. This reflects a higher friction for non-ionic and cationic polymers compared to the anionic polymer.

\begin{figure}
    \centering
    \includegraphics[width=0.99\linewidth]{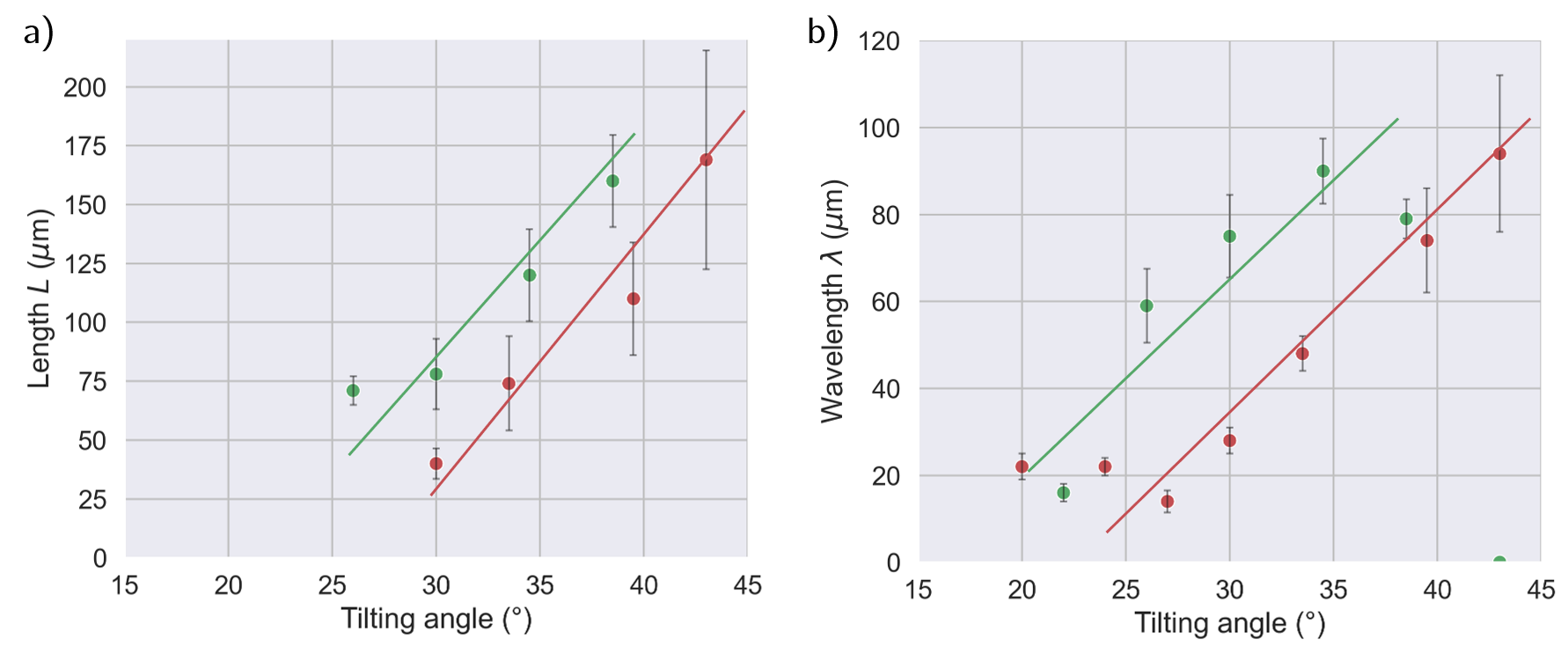}
    \caption{(a) Length $L$ and (b) wavelength $\lambda$ of the filaments as a function of tilting angle for the non-ionic (green) and cationic (red) polymer. Error bars correspond to standard deviation.}
    \label{fig:filaments_charac}
\end{figure}

To further analyze the characteristics of the filaments as a function of tilting angle, we measured the average filament length $L$ and average distance between filaments $\lambda$. For non-ionic and cationic polymers, $L$ and $\lambda$ increased with the tilting angle, reaching up to 175 $\mu$m and 90 $\mu$m, respectively. The wavelength $\lambda$ trend is consistent with previous studies, reported for a surface pulled from a liquid bath by Sharma \textit{et al.}\cite{sharma_newtonian_2025} and blade coating by Deblais \textit{et al.}\cite{deblais_taming_2016}. Overall, our results demonstrate that viscoelastic sliding drops exhibit receding contact line instabilities, leading to regularly spaced filaments. Filament formation depends on drop velocity and polymer charge. 

\subsection{Hydrodynamics of viscoelastic wetting}

The presence of polymers may alter drop motion and shape via two effects: either hydrodynamics (polymers rendering the solution viscoelastic and shear-thinning) or polymer-surface interaction. Let us first consider the hydrodynamics of viscoelastic wetting. For a Newtonian fluid, it is well-known from Cox-Voinov theory that the advancing (receding) contact angle increases (decreases) as the capillary number increases due to increased viscous stress. Accounting for hydrodynamics considerations, Kansal \textit{et al.} \cite{kansal_viscoelastic_2024,kansal_viscoelastic_2025} and Bartolo \textit{et al.} \cite{bartolo_dynamics_2007} showed that normal stresses arising from viscoelasticity only weakly affect the advancing contact line and contact angle but have strong effect on the receding contact line. It leads to a decrease in the receding contact angle compared to a Newtonian fluid. Regarding the advancing contact line, we indeed find no difference with that of pure water, neither with regard to the contact angle nor contact line smoothness. Moreover, we do observe a decrease in receding contact angle as we increase tilting angle and therefore capillary number. This increase is less than 10° for the anionic polymer drop, whereas it is more than 30° for the cationic and non-ionic polymer drops. Yet, the relaxation times of the solutions are 0.7\;s for the anionic, 0.19\;s for the cationic, and 0.04\;s for the non-ionic polymer drop. From hydrodynamics considerations, we would therefore expect a stronger effect for the anionic polymer drop. As this is not observed experimentally, we conclude that polymer-surface interactions play an important role. 

Regarding the triggering of the instability and filament formation, Sharma \textit{et al.} \cite{sharma_newtonian_2025} theoretically show that a receding liquid film becomes unstable below a critical thickness due to an interplay between Van der Waals and capillary forces. In Newtonian fluids, this leads to the emission of satellite droplets. In viscoelastic fluids, it results in filament formation: the elastic component (i.e., non-zero relaxation time) delays satellite droplet emission and make filament formation possible. However, they did not investigate how the filament characteristics (length, wavelength) depend on the solution. Yet, we observed significant differences in drop motion and contact line morphology depending on the polymer's charge. We now discuss the origin of these differences.

\subsection{Effect of polymer charge}

\begin{figure}
    \centering
    \includegraphics[width=0.99\linewidth]{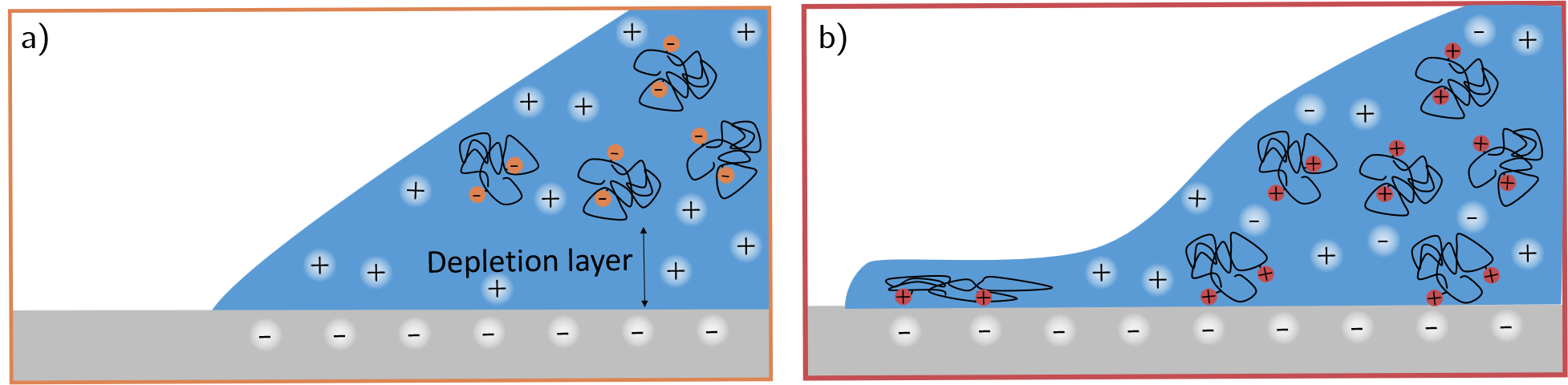}
    \caption{Schematic of the rear side of a sliding drop representing polymer adsorption at the solid-liquid interface depending on its charge: (a) anionic and (b) cationic polyelectrolyte.}
    \label{fig:schematics_polymer_adsorption}
\end{figure}

When discussing the interactions between the different types of polymers and the surface, we take into account that Teflon AF surfaces acquire a negative charge when exposed to water at neutral pH. Dissolved anionic polyelectrolytes are depleted near a negatively charge wall, as has been shown theoretically \cite{de_gennes_polymer_1981} and experimentally\cite{cross_wall_2018,guyard_near-surface_2021}; i.e. a layer with lower polymer concentration is formed (Fig.\;\ref{fig:schematics_polymer_adsorption}). This is due to electrostatic repulsion between the surface and the polymer. The depletion layer has been measured for a similar polymer to be about 30\;nm thick.\cite{guyard_near-surface_2021}
% Add reference OK
This is explained theoretically by balancing the monomer-wall interactions, which are repulsive in this case, with the osmotic pressure, which tends to homogenize polymer concentration. In terms of wetting properties, we therefore expect the static advancing and receding contact angles to be close to those of water, as well as a reduced viscoelasticity in the depletion layer. 

In contrast to the anionic polymer, the cationic polymers adsorb on the surface\cite{de_gennes_polymer_1981}, forming an arrested and stagnant
% maybe also add the term "stagnant" OK
region. We suggest that polymer adsorption reduces the solid-liquid surface energy, rendering the surface more hydrophilic, which leads to a higher drop friction and contact angle hysteresis. 

In both anionic and cationic cases, we observe protrusions in the contact line. It shows that the liquid layer at the receding contact line has a thickness lower than the critical thickness determined in \cite{sharma_newtonian_2025}. For the anionic case, we attribute the fact that no filaments are formed to the reduced elasticity in the depletion layer: the instability occurs but the oscillations are damped faster. In contrast, for the cationic case, polymer adsorption and solution elasticity favors filament formation. 

The formation of a depletion layer or of an absorbed layer also impact the liquid flow at the surface. For the anionic polymer a slip length is added while for cationic polymer the slip plane is shifted into the drop. However, these lengths are of the order of magnitude of the polymer radius of gyration. Compared to the size of the drop, this is negligible. We discuss later the evolution of the filament length over time as well as the deposition of polymer on the surface.

\begin{figure}
    \centering
    \includegraphics[width=0.7\linewidth]{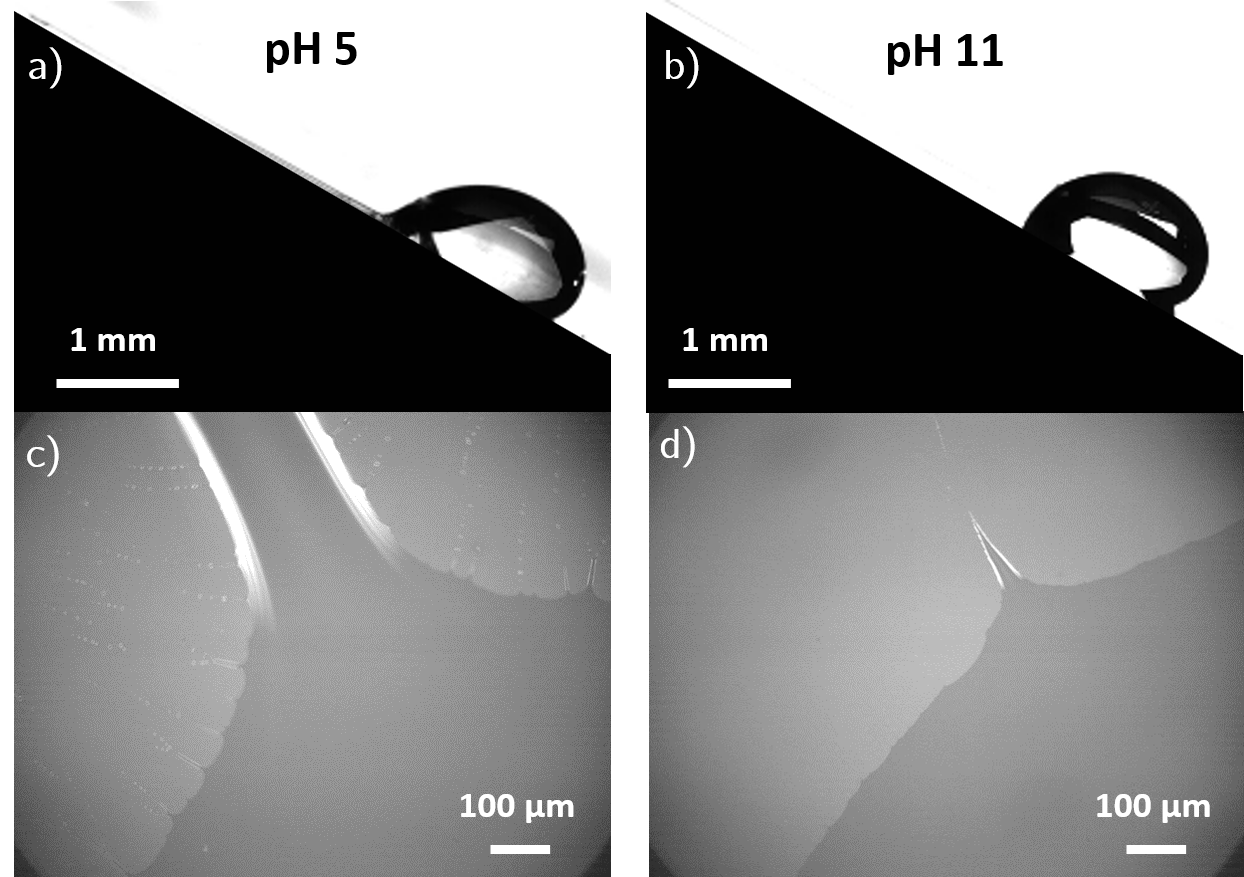}
    \caption{Side (a-b) and bottom (c-d) view of the rear side of a moving drop at pH 5 and 11.}
    \label{fig:pH}
\end{figure}

For the non-ionic polymer the situation is more complex. Polyacrylamide is considered to be neutral. Adsorption to the solid-liquid interface should not be enhanced by electrostatic interactions. Previous works have already demonstrated adsorption of polyacrylamide on silica surfaces\cite{pefferkorn_polyacrylamide_1999}. It had been explained by the formation of hydrogen bonds that, however, we do not expect on Teflon AF. To better understand an adsorption similar to cationic polymer, we study the effect of pH on the formation of filaments. Note that the initial solution has been made by mixing milliQ-water (at neutral or slightly lower pH due to CO$_2$ uptake) and the dry polymer. When dissolving polyacrylamide in water, we measured a pH of $\approx \!8$. This observation indicates that polyacrylamide binds some hydronium ions, becomes protonated, and charges slightly positive. To validate this hypothesis, we performed experiments at pH 5 and pH 11. Changing the pH affects both the charge of the polymer and the Teflon AF surface. Nevertheless, for the range of pH tested, Teflon AF remains negative\cite{zimmermann_electrokinetic_2001}. The new solutions are prepared by adding respectively hydrochloric acid and sodium hydroxide to the initial solution. The volume added is small so that changes in polymer concentration remained below 5\%. From the side view (Fig.\;\ref{fig:pH}a-b), we observe a difference in the receding contact angle of about 30°: 56$\pm$3° at pH 5 compared to 86°$\pm$3° at pH 11. From the bottom view (Fig.\;\ref{fig:pH}c-d), we observe the formation of filaments and the deposition of microscopic droplets at low pH, that we do not observe at high pH. This confirms a stronger adsorption at low pH, for which we expect more amine groups to be protonated, than at high pH, for which we expect the polymer to be neutral. 

At low pH, we also observe the formation of a tail of 2\;mm length at the rear of the drop. This tail exists also at high pH but is only 100\;$\mu$m long. This observation again indicates weaker adsorption. This tail is reminiscent of the pearling of sliding drops above a critical capillary number, studied in \cite{podgorski_corners_2001}. However, in our case, the transition to pearling is due to a decrease in the receding contact angle rather than to an increase in speed. These results highlight the critical role played by polymer-surface interaction in triggering instability and filament formation. It is consistent with the idea that adsorbed polymers decrease the contact angle, thereby modifying the solid-liquid surface energy. 

\subsection{Polymer deposition}

\begin{figure}
    \centering
    \includegraphics[width=0.99\linewidth]{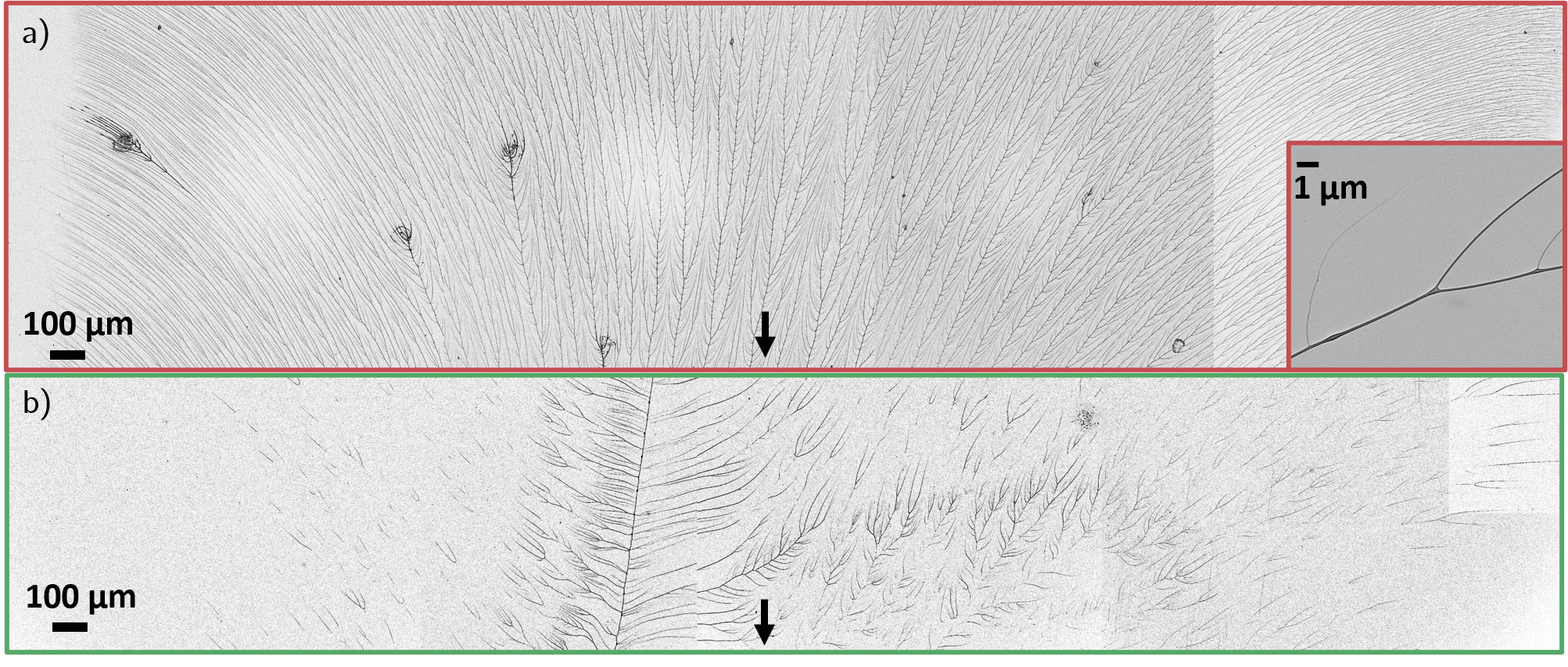}
    \caption{Cationic (a) and non-ionic (b) polymer deposition observed by SEM after one drop sliding over the Teflon AF surface. Original SEM images have been analyzed using edge detection for better contrast. Black arrows indicate sliding direction.}
    \label{fig:deposition}
\end{figure}

In Fig.\;\ref{fig:images_filaments}, we also observe the breakup of the filaments in microscopic droplets. This is due to the Rayleigh-Plateau instability\cite{lindner_viscoelastic_2009}. But the microdroplets are connected by a thinner filament. These filaments and microdroplets further dry and we expect it to lead to the deposition of polymer on the surface. 

To further verify the deposition of polymer, we image the surface across the whole drop width using Scanning Electron Microscopy (SEM) after the sliding of a single drop (Fig.\;\ref{fig:deposition}). 
For the drop containing cationic polymer, we observed the deposition of polymer in the form of filaments with a well defined spacing in the direction of motion. Filaments are oriented with an angle of $\approx 45$° compared to the direction of motion. This reflects the different relative motion of the contact line on the side of the drop. In the case of drops with non-ionic polymer, we also observe deposition of polymer but it is more scarce and we do not observe a pattern of continuous filaments. Most of the deposition occurs along the tail observed at the rear of the drop. 
We attribute the deposition of polymer to its adsorption at the solid-liquid interface. The polymer may directly adsorb at the solid-liquid interface or first move to the air-liquid interface and later adsorb at the receding contact line\cite{https://doi.org/10.1002/adma.202420263}.
For drops with anionic polymer, SEM images did not reveal any deposition except close occasional dust particles, which might have locally pinned the droplet (see SI, Fig.\;S2). Using SEM on Teflon AF surfaces the resolution is limited and we can not resolve single chains. Therefore, we cannot exclude that deposition occurs at even smaller scale. However, these results show that adsorption is stronger for the cationic and non-ionic polymers than for the anionic polymer. Additional experiments with pure glycerol droplets sliding on Teflon AF further highlight the role of polymer deposition. These droplets showed no filament formation at the rear (see SI, Fig.\;S3). Since glycerol has no affinity for Teflon AF surfaces, the contact line remains fully mobile, the receding rim stays smooth, and the instabilities required for filament formation never develop.

% We should consider the deposition paper of Xiaoteng Zhou (Adv. Mater. 2025, 2420263) and also discus alternative pathways of the polymer to be deposited to the solid surface. We only discuss the direct adsorption at the solid-liquid interface. However, the polymer may also adsorb to the liquid surface (see Fig. 3a in Xiaotengs paper, path 3) or the electric potential of the drop enhance deposition at the rear of the drop (Fig. 3a, path 1).
% Léa: These drops do not deposit much charges so I would say that path 1 is unlikely.
% \cite{https://doi.org/10.1002/adma.202420263} OK

\subsection{PDMS-coated glass}

The aforementioned results are obtained for drops sliding on Teflon AF-coated glass slides. 
%Kaloian: it seems to simple to me simply "other surface" May be you can contrast rigid Teflon vs. mobile, liquid like PDMS OK
To test the generality of the results to other surfaces, we carried out experiments with drops sliding on glass slides coated with PDMS brushes for a tilting angle of 30 °. Like Teflon AF surfaces, PDMS-coated slides charge negatively in solution but, in contrast to rigid Teflon AF surfaces, PDMS-coated surfaces exhibit a liquid-like behavior\cite{sbeih_influence_2024}. We find that drops are slower on PDMS surfaces: 9\;mm/s in average for the anionic polymer drop (against 15\;mm/s on Teflon AF) and 8\;mm/s for the non-ionic polymer drop (instead of 14\;mm/s on Teflon AF). For the cationic polymer, the drop remains pinned and does not slide at the macroscopic scale and on the timescale of a few tens of seconds, even when increasing the tilt angles to 90°. 
A possible explanation for the slower motion on PDMS-coated surfaces is that the droplet deforms the PDMS layer, leading to the formation of a capillary ridge. In contrast to a rigid substrate, this introduces an additional mechanism of energy dissipation, namely, viscoelastic dissipation within the PDMS layer\cite{https://doi.org/10.1002/adma.202311470}.
% Maybe mention that this is the case on a macroscopic scale and the time scale of several seconds or minutes. It could be that there is a slow motion that goes unnoticed. Stefan Karptschka and Jacco Snoeijer deteced such a slow motion on PDMS layer, although these were thick. OK

\begin{figure}
    \centering
    \includegraphics[width=0.9\linewidth]{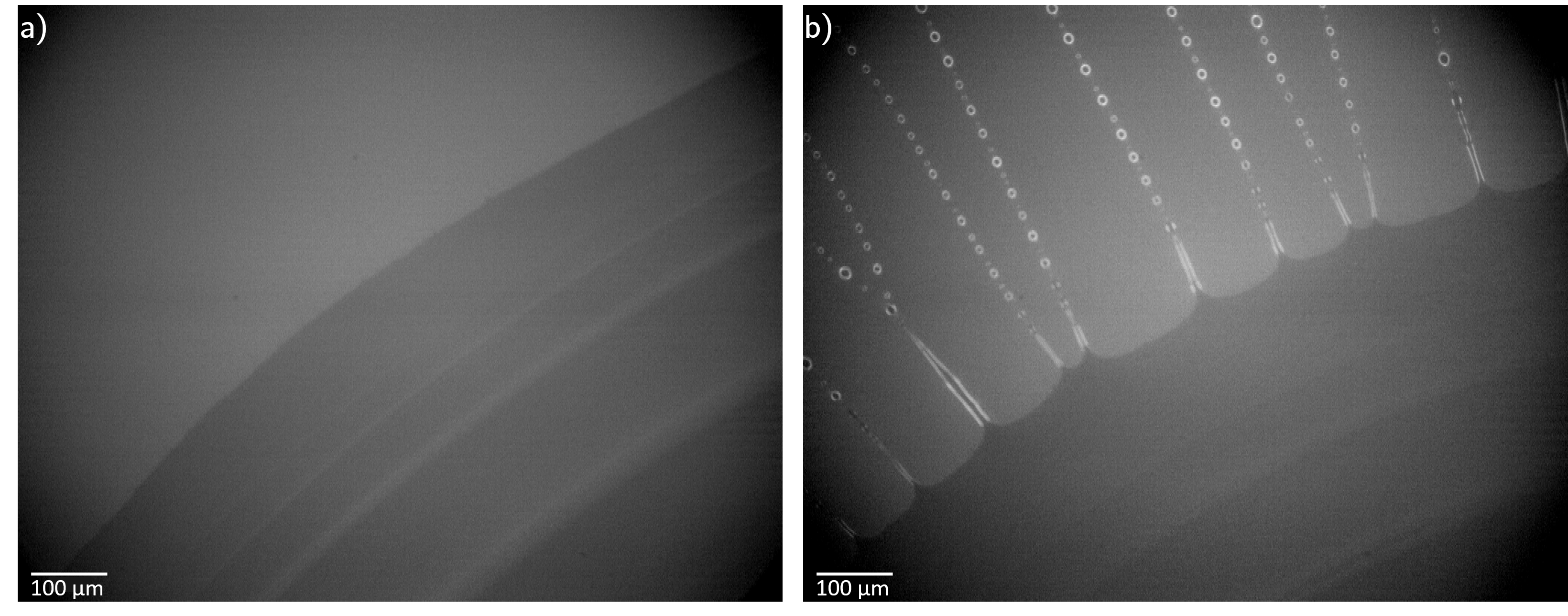}
    \caption{Receding contact line on PDMS for a) anionic and b) non-ionic polymer drops imaged by the inverted reflection microscope.}
    % For my taste the scale bar and "100 µm" could be slightly larger. OK
    \label{fig:pdms}
\end{figure}

This higher affinity for PDMS is also reflected in the contact angles: the advancing contact angle decreases from 123° and 117° on Teflon AF to 110° and 111° on PDMS, respectively, for the anionic and non-ionic polymers. Similarly, the receding contact angle decreases from 89° and 63° on Teflon AF to 74° and 58° on PDMS, respectively, for the anionic and non-ionic polymers.
% "We hypotheize that this difference is due to the entanglement between PDMS brushes and polymer present in the drops." Here I disagrre, PDMS and our three polymers would not mix. 
% Léa: Even the cationic one?
% I don't think so. Polymers harldy mix and the higher the molecular weight the less they mix. Silicones like themselves and typically do not mix with hydrocarbon based polymers. But I must admit, Moreover, the PDMS chains are not very long so that entanglement should not be a major factor for them.   OK
At the receding contact line of the two other polymers (Fig.\;\ref{fig:pdms}), we find similar qualitative shape: small deformation with few filaments for the anionic polymer drop and filaments of a $\approx\!100\;\mu$m for the non-ionic polymer drop. Therefore, although the macroscopic behavior depends on the substrate, we find that the shape of the contact line mostly depends on polymer charge. 
% Kaloian: you may like to state somewhere what is the charge of the PDMS coating vs Teflon coating OK
These results confirm the major role played by polymer-surface interaction in the formation of filaments. Moreover, the bright fringes observed in Fig.\;\ref{fig:pdms} are interferences fringes, which are reminiscent of the decrease in receding contact angle.

%How do we explain the slower motion of the drops on the PDMS coated surfaces? Viscoelastic deformation and ridge formation with the resulting viscous dissipation in the polymer layer? We saw a factor of two slower motion also for water drops sliding down PDMS-coated layers as compared to Teflon, see e.g. Zhou et al., Adv. Materials. 2024, 2311470, DOI: 10.1002/adma.202311470. I guess you can also cite Comparing methods for preparing slippery liquid-like polydimethylsiloxane coatings, I. J. Gresham, H. Barrio-Zhang, J. H. Cho, B. Khatir, G. G. Wells, K. Golovin, et al., Nature Protocols 2025. \cite{gresham_comparing_2025} OK

\subsection{Filament dynamics }

To understand the role of viscoelasticity in the filament formation, we estimated the magnitude of capillary, viscous and elastic stresses for the case of the cationic polymer. In this case, sliding drops of diameter $D_0 \approx$ 4~mm move at typical velocities $U\approx 1$\;mm/s. 
% To me it seems the velocities were more like 10 mm/s. At least the ranged from few mm/s to up to 60 mm/s.
% Which polymer are you talking about in the next sentence? I thought the filaments depend on the polymer. 
At the bulk scale, the characteristic shear rate of $\dot{\gamma}\approx U/D_0= 0.25~$s$^{-1}$, which matching this shear rate to the rheological curve of the cationic polymer at 2500 ppm concentration (see SI, Fig. S1) shows that the effective viscosity is quite close to the zero-shear value ($\eta_{eff}\!\approx\!0.4$\;Pa\;s). 
% Is the following estimation realistic? The shear rate for the drop is OK, because the top side of the drop is moving with a speed U (in fact it is probably even moving faster). But the top side of a filament is not moving, is it? At least it is not moving with a speed U. Please reconsider. My impression is that the following two sentences are not essential

% In the following part I am not expert enough in rheology. So here I rely on you. Please double check.  
However, filament thinning is governed by uniaxial extensional flow, and the relevant resistance is set by the extensional viscosity $\eta_E$, not the shear viscosity. For a viscoelastic liquid under uniaxial extension, the extensional stress~$\sigma_{el}$ is~\cite{mckinley2005visco}:
\begin{equation}
    \sigma_{el}=\tau_{zz}-\tau_{rr}=\eta_E  \dot{\epsilon}\,,
\end{equation}
where $\tau_{zz}$ and $\tau_{rr}$ correspond to the normal and radial stresses along the filament and $\dot{\epsilon}$ is the extensional rate from Dripping onto surface measurements (DoS):
% What does "DoS" stand for?
\begin{equation}
    \dot{\epsilon}=-\frac{2}{R(t)}\frac{dR}{dt}\,,
\end{equation}
Here, $R(t)$ corresponds to the minimum filament radius~($d_{min}/2$) over time obtained from the DoS high-speed recordings, assuming a cylindrical shape. The extensional viscosity can be expressed from the capillary balance as
% Kaloian: Please define this abbreviation on first appearance in the Methods on page 3
\begin{equation}
    \eta_E=-\frac{\sigma}{2\,dR(t)/dt}\,.
\end{equation}

Our DoS measurements shows a maximum of $\eta_E\approx 260$~Pa s and $\dot{\epsilon}=4.5$ (Fig.\;9), giving an elastic stress $\sigma_{el}\approx 1$~kPa. At the initial filament width $e_0=15\;\mu$m the capillary stress is $\sigma_{cap}=2\sigma/h = 9.6$~kPa. This is about one order of magnitude larger than $\sigma_{el}$. Despite such difference, elastic stresses are sufficient to sustain filaments and slow thinning, allowing the formation of multiple beads separated by finite tails/filaments (Fig.\;8). As thinning proceeds $\sigma_{cap}$ increases with $1/h$, eventually overcoming $\sigma_{el}$ to induce breakup. 
\begin{figure}
    \centering
    \includegraphics[width=\textwidth]{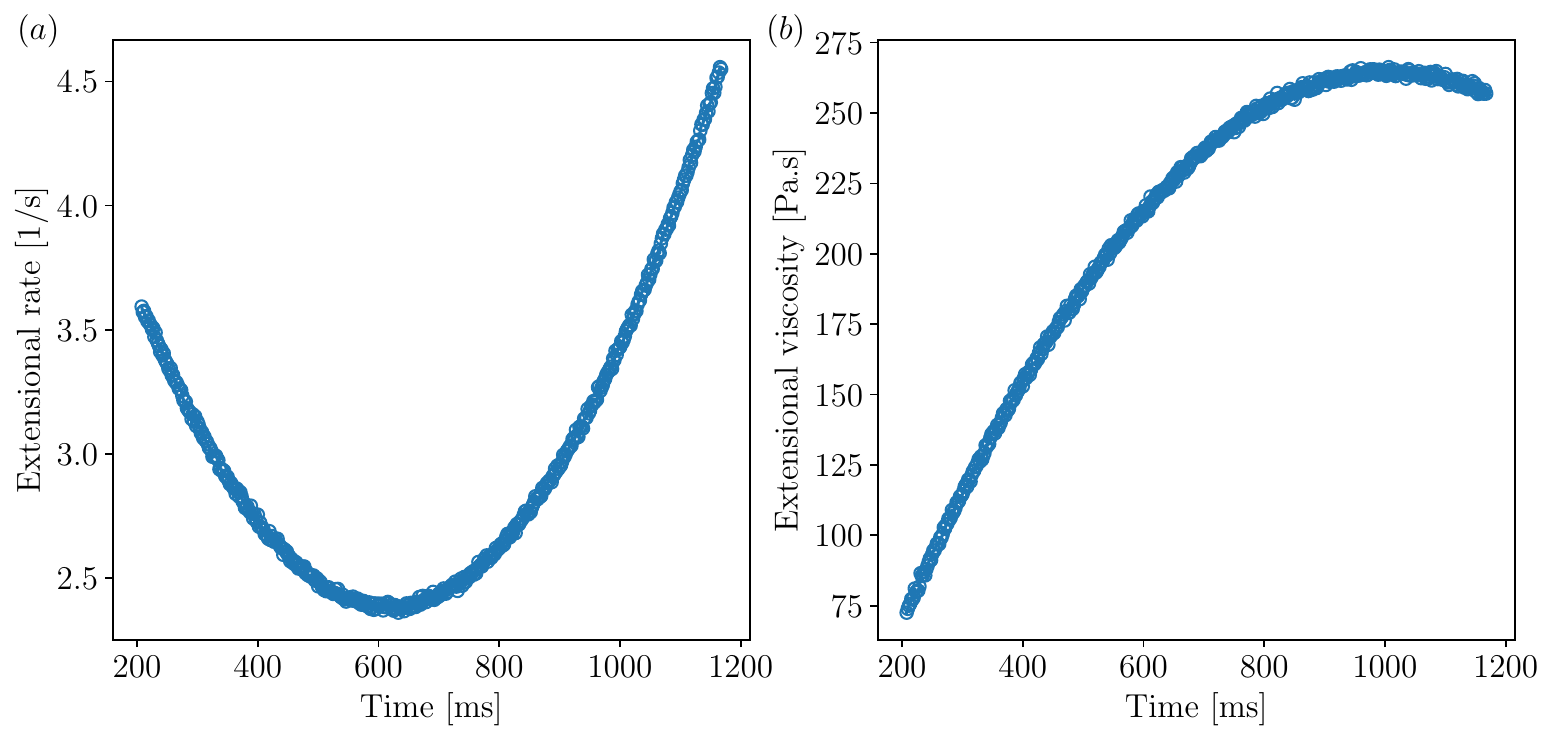}
    \caption{Extensional rate (a) and (b) extensional viscosity for the anionic polymer.}
    % Please give the values for the right vertical axis in e.g. 36 rather than 3.6 times 10^1.
    % Kaloian: you may refer to DoS in the figure caption
    \label{fig:placeholder}
\end{figure}

Viscous stresses can be estimated as $\sigma_{vis}=\eta_{eff}\dot{\gamma}_{fil}=2.6$ Pa, which are quite small in comparison with $\sigma_{el}$ and $\sigma_{cap}$. Therefore, viscous stresses are negligible as instantaneous tensile resistance strength. Their role is rather dissipative, damping the growth of capillary perturbations. The bead on a tail morphology is thus a result of a Rayleigh-Plateau instability modified by the viscoelasticity, where capillarity drives the process elasticity sustain filament formation and delays breakup, while viscosity governs viscous dissipation. Note that the filaments observed in the “tail on a bead” instability are not stable. Following the argument of Xu et al.~\cite{xu_sliding_2016,xu_viscoelastic_2018}, we compare the breakup time of a viscoelastic filament, $\tau_b \sim \eta_Ed/\sigma$, with a convective timescale $\tau_c=l/U$, where $l$ is a characteristic length along the tail. Taking $L=w$, the experimentally observed bead spacing, and using $d\approx w/2$, we obtain a critical extensional viscosity for the filament survival $\eta_E \sim 2\sigma/U\approx 144$ Pa s. This value is consistent with our observations: the extensional viscosity as determined from DoS is in the same order of magnitude that this threshold, and the filaments, although they clearly form, break up rapidly after only a few bead spacings. In contrast to previous studies on micropillared superhydrophobic surfaces, where the bead spacing is constrained by the pillar pitch and filaments are anchored on pinned droplets~\cite{xu_sliding_2016, xu_viscoelastic_2018}, the spacing in our experiments is not imposed by surface topography (PDMS and Teflon AF are smoother). Instead, it reflects the intrinsic wavelength selected by the Rayleigh–Plateau instability, whose growth is delayed by viscoelasticity but not suppressed. 

\section{Conclusion}
Our study demonstrates that viscoelastic sliding drops exhibit receding contact line instabilities, which lead to filament formation. The onset and extent of this instability depend critically on the charge of the polymer: cationic and non-ionic polymers promote the formation of long filaments, whereas anionic polymers show minimal deformation of the contact line. 
This behavior can be attributed to electrostatic interactions at the solid–liquid interface, leading to polymer adsorption in the case of cationic systems and depletion for anionic ones, and a modification of the wetting properties of the fluid on the surface. Notably, the polymer initially classified as non-ionic was found to be partially protonated and to adsorb onto the surface, behaving similarly to cationic polymers.
% Could you rephrase the last sentence? I am not yet 100% happy because the adsorption behavior and the states of protonation are not obviously linked. Maybe add another sentence saying that the non-ionic polymer is in fact not neutral but is partially protonated and as a result also adsorbs. OK
% Another preoblem of this sentence is thwe "dynamics". It leaves the impression that not the adsotprion itself plays a crucial role, but the specific dynamics of it. However, in the manuscript we do not address dynamics of adsorption, do we? We do not discuss, how long it takes for the polymer to adsorb. OK
% Kaloian: I think it may be nice to mention electrostatic interaction here
These findings emphasize the interplay between rheology, surface interactions, and drop dynamics, with potential applications in coating, printing, and surface engineering.

\section{Supplementary Material}

See the supplementary material for details regarding the solutions' rheology (flow curves, Fig. S1), SEM image following anionic droplet sliding (Fig. S2), and the sliding of glycerol drops (Fig. S3).

\begin{acknowledgments}
The authors would like to thank Andreas Hanewald and Gunnar Glasser respectively for rheology and SEM measurements, as well as Lin Jian for providing the PDMS-coated samples. We also thank PolyChemie GmbH for providing the polymers used in the study. This work was supported by the European Research Council (ERC) under the European Union’s Horizon 2020 research and innovation program (Grant agreement No. 883631) (H.-J. B.) and the Consolidator Grant INTER-ET (grant agreement no. ERC-CoG-2024-101171358) (D. D., O. T.).
\end{acknowledgments}

\section{Author Declarations}
The authors have no conflicts to disclose.

\section{Data availability}
The data that support the findings of this study are available from the corresponding author upon reasonable request. 

\bibliography{biblio_visco_elastic_drops}

\end{document}